# Design of an Ultra-Efficient Reversible Full Adder-Subtractor in Quantum-dot Cellular Automata

Elham Taherkhani[1], Mohammad Hossein Moaiyeri[1,2]*and Shaahin Angizi[3]

[1] Nanotechnology and Quantum Computing Lab, Shahid Beheshti University, G. C., Tehran, 1983963113, Iran
[2] Department of Electrical Engineering, Shahid Beheshti University, G. C., 1983963113, Tehran, Iran
[3] Department of Electrical and Computer Engineering, University of Central Florida, Orlando, FL 32816, USA
* Corresponding Author (E-mail: h_moaiyeri@sbu.ac.ir; Tel.: +982129904105)

**Abstract**
By the progressive scaling of the feature size and power consumption in VLSI chips the part of energy dissipated due to information loss in irreversible computations will become a serious limitation in the near future. Quantum-dot cellular automata (QCA) is an emerging nanotechnology with extremely low energy dissipation which facilitates new computation paradigms such as reversible computing. In this paper a novel reversible full adder-subtractor circuit based on QCA is proposed. Our proposed design is implemented using only one layer and does not require any rotated cells which significantly improves the manufacturability of the design. In addition, it improves the cell count, area and total energy dissipation by almost 45% and 50% and 48%, respectively, as compared to the existing QCA-based single-layer and multilayer reversible full adders.

## 1. Introduction

Power consumption is indeed the main concern in recent VLSI circuits. Due to the ever-increasing demands for portable electronic systems and the disproportionate growth of the semiconductor and battery manufacturing industries, increasing the operational time between each battery charge has become very curtail. In addition, the disproportionate scaling of the size of transistors and power supply voltage has led to many shortcomings such as high leakage currents and high power density, creating hot spots in CMOS chips. As a result, different computational paradigms using emerging nanotechnologies, addressing drastically small size and low power dissipation, should be considered. Reversible computing is one these computational paradigms which is realized by setting up a one-to-one mapping between the input and output states of the circuit [1-3].

In 1973 Bennett proved that it is possible to eliminate power dissipation in a logic circuit if the circuit includes only reversible logic gates. Moreover, in 1961 Landauer demonstrated that each bit of information loss in an irreversible computation leads to $k_BTln2$ joules dissipation of heat energy, where $k_B$ is Boltzmann constant and $T$ is temperature. As a result, the energy required for a binary transition at room temperature (T=300 K) is at least 0.017 eV. However, if a computation is carried out in a reversible manner, this amount of energy would not inevitably dissipated and energy consumption would be beyond the $k_BTln2$ limit. Although the power dissipation of the current CMOS circuits is much higher than $k_BTln2$, in the near future with ultimate scaling of feature size and power consumption based on emerging nanotechnologies, this theoretical limit may become a major restriction [3].

Utilizing reversible computing in nanotechnology and quantum computing has become a quite interesting subject in recent years. Reversible computing, as a new paradigm for reducing physical entropy gain, is based on physical operations, which are both logically and thermodynamically reversible [4]. One the most attractive and promising nanotechnologies is quantum-dot cellular automata (QCA) which operates based on the new physical phenomena such as Coulombic interactions [1,5].

In QCA, two distinct situations of electrons inside the cell determine the '0' and '1' logics as shown in Fig. 1(a). The electrons can quantum-mechanically tunnel between the dots though tunneling junctions and fix either with cell polarization *P*=-1 (logic '0') or in *P*=1 (logic '1') as illustrated in Fig. 1 (a). This polarity can be defined by (1), where $\rho_i$ is the probability of presence of electron in quantum-dot *i* [6].

$$P = \frac{\rho_1 + \rho_3 - \rho_2 + \rho_4}{\rho_1 + \rho_3 + \rho_2 + \rho_4} \qquad (1)$$

Hence, when a bit changes from '0' to '1' and vice versa, no actual charging and discharging of capacitors occur as in CMOS. Moreover, transmission of logic from one QCA cell to another is performed by the interaction of electrons in neighbouring cells and there is no current flow between QCA cells. Accordingly, in QCA, there is no energy dissipation during the state transition and propagation. As a result, QCA dissipates extremely lower power as compared to the CMOS technology. Also, the part of energy dissipated due to information lost in irreversible computations becomes more important. It has been demonstrated that the energy dissipated per switching event in a reversible QCA circuit can be significantly less than $k_BTln2$ due to clocked information-preserving system. As a result, implementing efficient reversible logic gates becomes possible using QCA [7].

The fundamental QCA gates are the inverter and the majority (MAJ) gates which are shown in Fig. 1 (b) and (c), respectively. The QCA inverter, which is usually realized in two different configurations, reverses the cell polarization and the output of the majority gate will be the majority if the inputs. A series of QCA cells creates a QCA wire in which the binary data propagates because of the Columbic interactions between the cells [8]. A four phase clocking system is utilized to provide the propagation of data through the logic circuits [9]. This clocking scheme including switch, hold, release and relax phases is illustrated in Fig. 2.

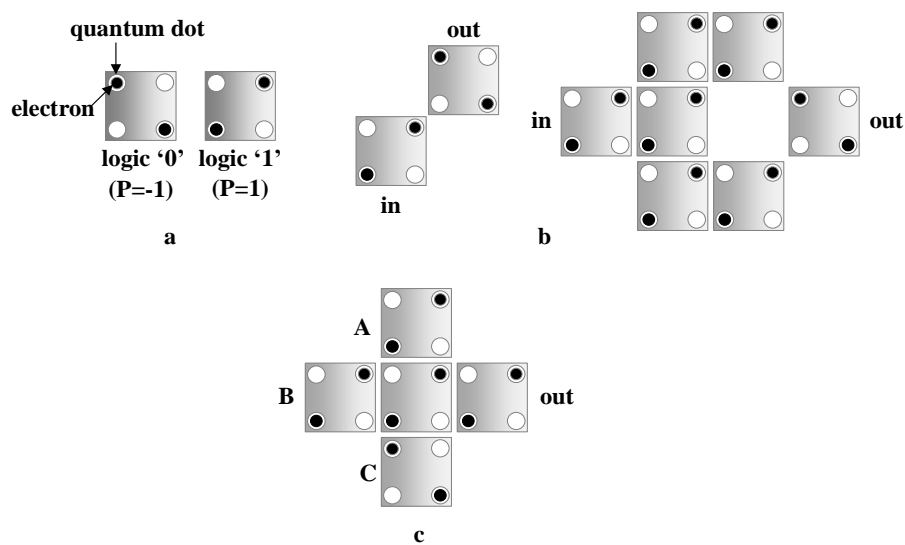

Fig. 1. QCA logic (a) QCA cells representing logic '0' and logic '1'(b) QCA simple inverter c QCA three-input majority

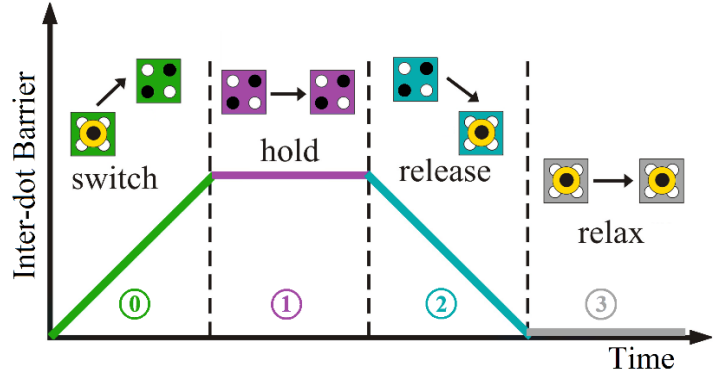
Fig. 2. Different clocking zones in QCA design

Arithmetic unit is one of the most power hungry sections in a processor and is the most probable location of hot-spots [10]. Full adder is the most important building block of arithmetic units and improving the efficiency of this circuit leads to improvement of the efficiency of the whole processor. Due to the importance of the reversible computation based on QCA some researches have already been done for designing QCA-based reversible logic gates as the building blocks of QCA-based reversible computational circuits such as full adder cells [1,3,11]. However, they lead to cell count and area redundancy for designing full adder circuit which is the most fundamental block of the arithmetic unit in each digital processor. In addition, they require rotated cells or multilayer wire crossings which lead to lower manufacturability and robustness.

In this paper, a novel efficient reversible adder-subtractor circuit in QCA nanotechnology is proposed. Our proposed design is implemented in one layer and has significantly lower number of cells, lower energy consumption and smaller area even in comparison with its multilayer counterparts.

## 2. Proposed Designs

The suggested reversible gate called RQG is shown in Fig. 3 (a). This reversible gate has three inputs X1, X2 and X3 and generates three outputs Y1, Y2 and Y3 according to (2) to (4). The RQG gate is fully reversible with a completely one-to-one mapping between the inputs and outputs which can be obtained from its truth table shown in Table 1.

$$Y_1 = MAJ(X_1, X_2, X_3) = X_1.X_2 + X_2.X_3 + X_1.X_3 \qquad (2)$$

$$Y_2 = MAJ(X_1', X_2, X_3)) = X_1'.X_2 + X_2.X_3 + X_1'.X_3 \qquad (3)$$

$$Y_3 = X_1 \oplus X_3 = X_1.X_3' + X_1'.X_3 \qquad (4)$$

The QCADesigner tool [12], as widely used QCA circuit design and verification, has been used to implement and verify all of the designs in this study. The QCA layout of this reversible gate is shown in Fig. 3 (b). As illustrated in the layout the RQG gate is composed of only two three-input majority gates and one simple 2-input exclusive-OR (XOR) gate. In this efficient QCA layout, a simple and dense two-input exclusive-OR (XOR) design [13] is utilized for realization of the third output ($Y_3$).

In addition, for wire crossing, the single layer method using different clock zones is used [14,15]. In general, wire crossing in QCA circuits can be performed by tree approaches using multilayer cells, rotated cells and different clock zones. The feasibility of fabricating multilayer QCA cells is still a questionable challenge. Moreover, fabricating rotated cells is quite difficult and require very accurate alignment during fabrication [16]. However, wire crossing using different clock phases is performed in one layer without needing any rotated cells. Hence, the layout of RQG is designed using only one layer of normal QCA cells.

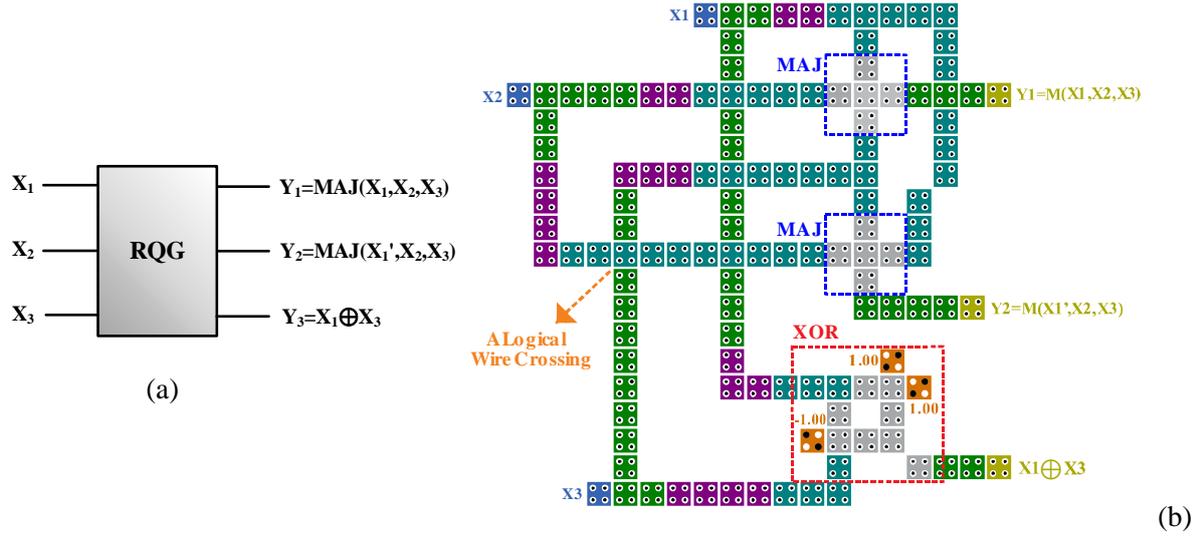

Fig. 3. The suggested reversible gate (a) Symbol view (b) Layout view

Table. 1. Truth table of the suggested reversible gate

| $X_1$ | $X_2$ | $X_3$ | $Y_1$ | $Y_2$ | $Y_3$ |
|---|---|---|---|---|---|
| 0 | 0 | 0 | 0 | 0 | 0 |
| 0 | 0 | 1 | 0 | 1 | 1 |
| 0 | 1 | 0 | 0 | 1 | 0 |
| 0 | 1 | 1 | 1 | 1 | 1 |
| 1 | 0 | 0 | 0 | 0 | 1 |
| 1 | 0 | 1 | 1 | 0 | 0 |
| 1 | 1 | 0 | 1 | 0 | 1 |
| 1 | 1 | 1 | 1 | 1 | 0 |

The RQG gate as a versatile reversible gate can be used to implement important logic circuits such full adder and full subtractor. By using an RQG and two Feynman reversible gates (FG), a novel reversible full adder-subtractor circuit with four inputs and four outputs is proposed which is shown in Fig. 4. It is worth noting that a Feynman gate with x and y inputs, provides x and x⊕y outputs [1]. In the proposed design, the first FG replicates the B input and the second one is used to create a three input XOR (A⊕B⊕C). The three main outputs of the circuit provides the full adder-subtractor as given in (5)-(7), where Sum and Diff are the sum and difference of the three inputs, respectively and Cout and Bout are the output carry and borrow, respectively. Furthermore, the forth output (A⊕C) is a garbage output. The truth table of the proposed reversible full adder-subtractor is shown in Table 2, which indicates the completely one-to-one mapping between the inputs and outputs.

$$C_{out} = MAJ\ (A,B,C) = A.B + B.C + A.C \tag{5}$$

$$B_{out} = MAJ\ (A',B,C) = A'.B + B.C + A'.C \tag{6}$$

$$Sum = Diff = A \oplus B \oplus C = A.B.C + A'.B'.C + A'.B.C' + A.B'.C' \tag{7}$$

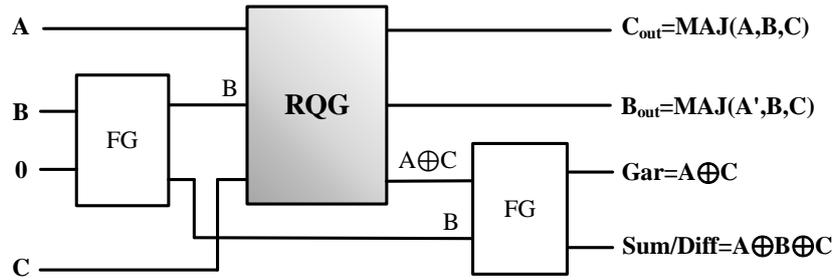

Fig. 4. Proposed reversible full adder-subtractor

The layout of the proposed reversible full adder-subtractor is shown in Fig. 5. In this efficient QCA layout an efficient QCA implementation for FG is suggested based on the dense XOR of [12] as shown in Fig. 5. In addition, the logical crossing method using different clock zones is used for wire crossing [14,15]. Accordingly, the layout of the proposed full adder-subtractor is designed using only one layer of normal QCA cells, which enhance the robustness and manufacturability of the proposed design.

Table. 2. Truth table of the proposed reversible full Adder-Subtractor

| A | B | 0 | C | $C_{out}$ | $B_{out}$ | Sum/Diff | Gar |
|---|---|---|---|---|---|---|---|
| 0 | 0 | 0 | 0 | 0 | 0 | 0 | 0 |
| 0 | 0 | 0 | 1 | 0 | 1 | 1 | 1 |
| 0 | 1 | 0 | 0 | 0 | 1 | 1 | 0 |
| 0 | 1 | 0 | 1 | 1 | 1 | 0 | 1 |
| 1 | 0 | 0 | 0 | 0 | 0 | 1 | 1 |
| 1 | 0 | 0 | 1 | 1 | 0 | 0 | 0 |
| 1 | 1 | 0 | 0 | 1 | 0 | 0 | 1 |
| 1 | 1 | 0 | 1 | 1 | 1 | 1 | 0 |

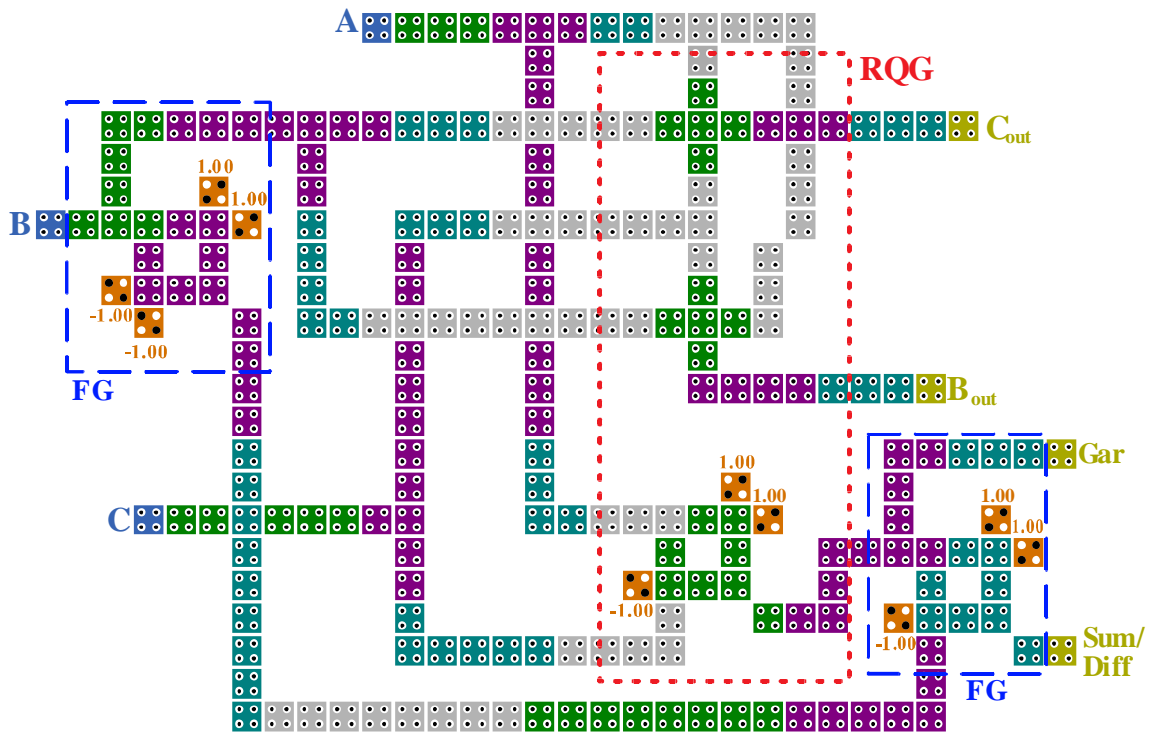

Fig. 5. The QCA implementation of the proposed full adder-subtractor

## 3. Performance evaluation

QCADesigner software as a popular simulation tool for QCA circuits is used for simulation and verification of then designs. All the simulation parameters and conditions are adopted as the default values in QCADesigner tool as follows: QCA cell size=18 nm, diameter of quantum dots=5 nm, number of samples=50,000, convergence tolerance=0.001, radius of effect=65 nm relative permittivity=12.9, clock low=3.8e-23 J, clock high=9.8e-22 J, clock amplitude factor=2.000, layer separation=11.5 nm and maximum iterations per sample=100. The simulation results, shown in Fig. 6 validate the functionality of the proposed reversible full adder-subtractor which utilizes the RQG gate as its main block.

Table 3 includes a comparison between our proposed full adder-subtractor design and the state-of-the-art existing reversible designs. The proposed reversible full adder-subtractor produces only one garbage output which is fewer than the other designs, while it has only one constant input like [3]. The sum of the garbage outputs and constant inputs is two for our proposed design, while it is three for the other designs. Using different clock zones for wire crossing makes the latency (number of clock cycles) of the proposed design slightly longer than [1] and [17], while it is considerably shorter than [3]. However, unlike the previous works, our proposed design is single-layer without any rotated cells which significantly enhance its robustness and manufacturability. Also, it performs both addition and subtraction operations. In addition, it has a considerably lower number of cells and much smaller area in comparison with both single-layer and multilayer designs.

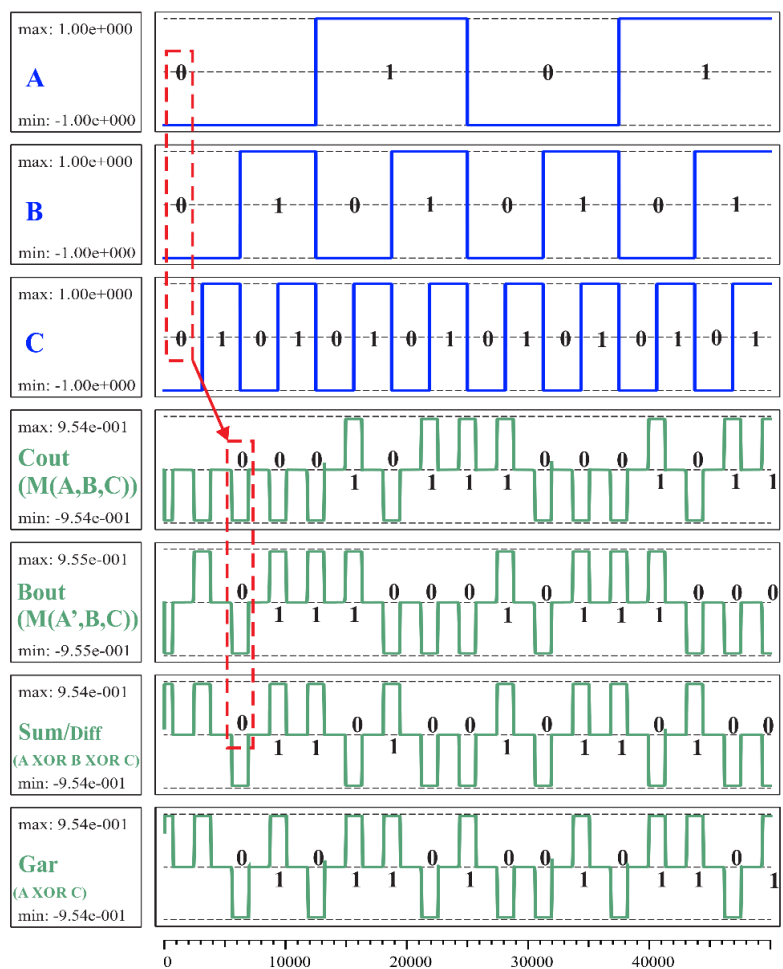

Fig. 6. Simulation results of the proposed full adder-subtractor

Table 3. Comparison of QCA reversible full adders

| Designs | Cells | Rotated Cells | Layers | Area ($\mu m^2$) | Latency (clock cycles) | Constant Inputs | Garbage Outputs |
|---|---|---|---|---|---|---|---|
| QCA1 [1] | 343 | 110 | 1 | 0.46 | 1.50 | 0 | 3 |
| QCA2 [1] | 356 | 105 | 1 | 0.47 | 1.50 | 0 | 3 |
| FG+RQCA [3] | 517 | 155 | 1 | 0.78 | 3.25 | 1 | 2 |
| RM [11] | 612 | 175 | 1 | 0.96 | 4.00 | 3 | 3 |
| [17] | 351 | 0 | 2 | 0.41 | 1.50 | 0 | 3 |
| **Proposed** | **228** | **0** | **1** | **0.28** | **1.75** | **1** | **1** |

QCAPro [18] as a valid energy estimator tool can be readily exploited to evaluate the leakage, switching, and total energy dissipations of the QCA circuits. Table 4 represents the energy dissipation analysis of the proposed QCA reversible full adder-subtractor based on RQG and recently reported counterparts. The simulations are performed in three distinct tunneling energy levels (0.5, 1, and 1.5 $E_k$) considering 2K as the operational temperature. Besides, Fig. 7 shows the power dissipation map of our design at 2K temperature and 1.5 $E_k$. It is worth pointing out that the darker the QCA cells are, the more energy dissipates in the circuit.

Table 4. The energy dissipation analysis of the proposed QCA reversible full adders.

| | Avg. leakage energy dissipation (eV) | | | Avg. switching energy dissipation (eV) | | | Total energy consumption (eV) | | |
|---|---|---|---|---|---|---|---|---|---|
| Designs | 0.5 $E_k$ | 1 $E_k$ | 1.5 $E_k$ | 0.5 $E_k$ | 1 $E_k$ | 1.5 $E_k$ | 0.5 $E_k$ | 1 $E_k$ | 1.5 $E_k$ |
| QCA1 [1] | 0.085 | 0.235 | 0.452 | 0.342 | 0.285 | 0.259 | 0.427 | 0.52 | 0.711 |
| QCA2 [1] | 0.093 | 0.241 | 0.475 | 0.348 | 0.301 | 0.292 | 0.441 | 0.542 | 0.767 |
| FG+RQCA [3] | 0.202 | 0.566 | 0.964 | 0.426 | 0.35 | 0.286 | 0.628 | 0.916 | 1.25 |
| RM [11] | 0.351 | 0.954 | 1.602 | 0.927 | 0.753 | 0.615 | 1.278 | 1.707 | 2.217 |
| [17] | 0.094 | 0.271 | 0.486 | 0.354 | 0.309 | 0.275 | 0.448 | 0.579 | 0.761 |
| **Proposed** | **0.068** | **0.209** | **0.376** | **0.269** | **0.236** | **0.203** | **0.337** | **0.445** | **0.579** |

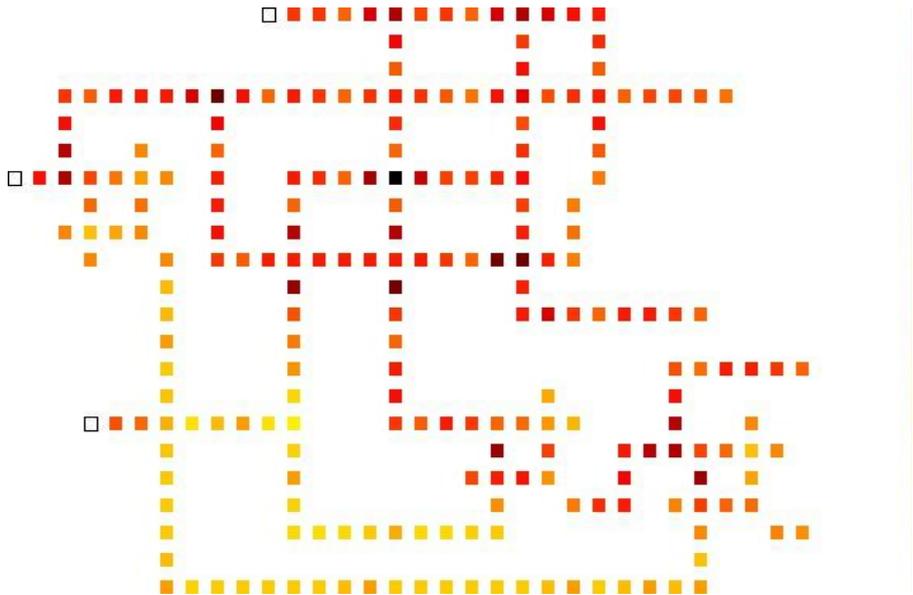

Fig. 7. Energy dissipation map for the proposed QCA reversible design at 2K temperature with 1.5 $E_k$.

The graphical diagram depicted in Fig. 8 shows the total energy consumption comparison results of the designs. These diagram can attest the energy efficiency of our proposed structure. According to Fig. 8, it can be easily perceived that our design has much less energy dissipation in different tunneling energy levels in comparison to the previously reported designs.

Appropriate cell arrangements along with not using rotated cells in the proposed design are the most important reasons for such minimized energy consumption. In an objective comparison of total energy dissipation (over all vector pairs) to the previously reported reversible QCA full adders listed in Table 4, our proposed design dissipates on average 48%, 48%, and 50% less energy in $0.5E_k$, $1E_k$, and $1.5E_k$ tunneling energy levels, respectively.

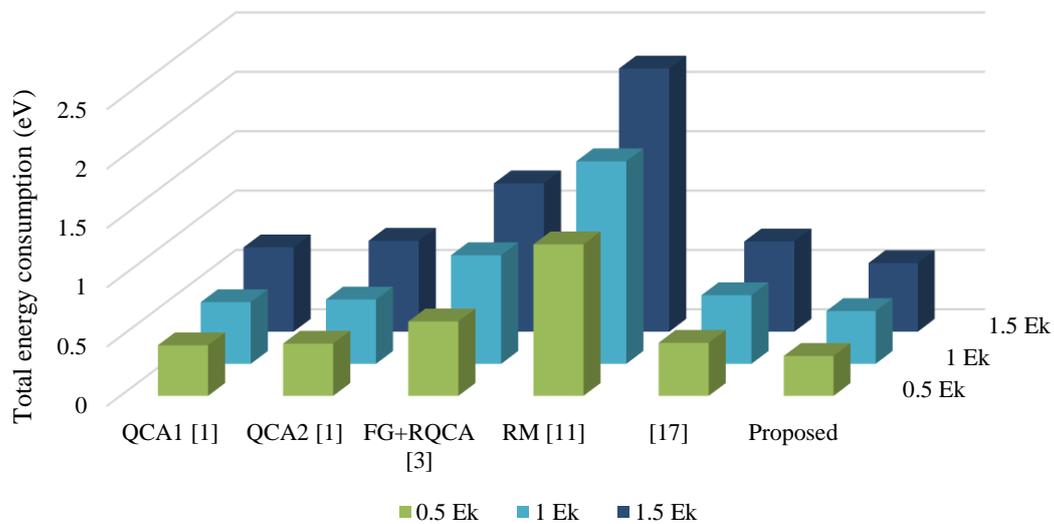

Fig. 8. The average energy dissipation of the QCA reversible design in diverse tunneling energy levels (overall vectors) at 2.0 K.

## 4. Conclusion

This latter presents a novel efficient QCA-based full adder-subtractor for reversible nanocomputing. While our proposed circuit provides both addition and subtraction operations and has a single-layer structure, it outperforms the previous single-layer and multilayer reversible full adders in terms of cell count and area. In addition, it does not require rotated cells and has a lower number of garbage outputs. Accordingly utilizing such an efficient reversible full adder-subtractor will certainly be a breakthrough in the field of computation.

## References

[1] X. Ma, J. Huang, C. Metra, and F. Lombardi, "Reversible and testable circuits for molecular QCA design", in Tehranipoor, M. (Ed.): "Emerging Nanotechnologies", Springer, US, 2008.
[2] M. Rezaeikhezeli, M. H. Moaiyeri and Ali Jalali, "Analysis of Crosstalk Effects for Multiwalled Carbon Nanotube Bundle Interconnects in Ternary Logic and Comparison with Cu interconnects", IEEE Transactions on Nanotechnology, Vol. 16, No. 1, 2017.
[3] B. Sen, M. Dutta, S. Some and B. K. Sikdar, "Realizing reversible computing in QCA framework resulting in efficient design of testable ALU", ACM Journal on Emerging Technologies in Computing Systems, vol. 11, no. 3, pp. 30:8–22, 2014.

[4] M. Haghparast and Ali Bolhassani, "On Design of Parity Preserving Reversible Adder Circuits", International Journal of Theoretical Physics, vol. 55, no. 12, pp. 5118-5135., 2016.

[5] N. G. Rao, P. C. Srikanth, and S. Preeta, "A Novel Quantum Dot Cellular Automata for 4-Bit Code Converters", Optik, vol. 127, pp. 4246-4249, 2016.

[6] C. S. Lent and P. D. Tougaw, "A device architecture for computing with quantum dots", in Proc. IEEE, vol. 85, no. 4, pp. 541-557, 1997.

[7] J. Timler and C. S. Lent, "Power gain and dissipation in quantum-dot cellular automata", Journal of Applied Physics, vol. 91, no. 2, pp. 823–831, 2002.

[8] E. Taher Karkaj, S. Rasouli Heikalabad, "Binary To Gray And Gray To Binary Converter In Quantum-Dot Cellular Automata", Optik, vol. 130, pp. 981-989, 2017.

[9] S. Sheikhfaal, S. Angizi, S. Sarmadi, M. H. Moaiyeri and S. Sayedsalehi, "Designing Efficient QCA Logical Circuits With Power Dissipation Analysis," Elsevier, Microelectronics Journal, Vol. 46, No. 6, pp. 462-471, 2015.

[10] S. Ghosh and K. Roy, "Exploring high-speed low-power hybrid arithmetic units at scaled supply and adaptive clock-stretching", Asia and South Pacific Design Automation Conf., pp. 635–640, 2008.

[11] B. Sen, M. Dutta, M. Goswami, and B. K. Sikdar, "Modular Design of testable reversible ALU by QCA multiplexer with increase in programmability", Microelectronics Journal, vol. 45, no. 11, pp. 1522-1532, 2014.

[12] K. Walus, T. J. Dysart, G. A. Jullien and R. A. Budiman, "QCADesigner: a rapid design and simulation tool for quantum-dot cellular automata", IEEE Transactions on Nanotechnology, vol. 3, no. 1, pp. 26–31, 2004.

[13] A. M. Chabi, A. Roohi, R. F. DeMara, S. Angizi, K. Navi, and H. Khademolhosseini, "Cost-efficient QCA reversible combinational circuits based on a new reversible gate", Computer Architecture and Digital Systems CSI Int. Symp., pp. 1–6, 2015.

[14] S. H. Shin, J. C. Jeon and K. Y. Yoo, "Wire-crossing technique on quantum-dot cellular automata", Next Generation Computer and Information Technology Int. Conf., pp. 52-57, 2013.

[15] F. Ahmad, G. M. Bhat, H. Khademolhosseini, S. Azimi, S. Angizi and K. Navi, "Towards single layer quantum-dot cellular automata adders based on explicit interaction of cells". Journal of Computational Science, vol. 16, pp. 8-15, 2016.

[16] C. Labrado and H. Thapliyal, "Design of adder and subtractor circuits in majority logic-based field-coupled QCA nanocomputing", Electronics Letters, vol. 52, no. 6, 464-466, 2016.

[17] Z. Mohammadi and M. Mohammadi, "Implementing a one-bit reversible full adder using quantum-dot cellular automata", Quantum Information Processing, vol. 13, no. 9, pp. 2127–2147, 2014.

[18] S. Srivastava, A. Asthana, S. Bhanja, and S. Sarkar, "QCAPro-an error-power estimation tool for QCA circuit design," in 2011 IEEE International Symposium of Circuits and Systems (ISCAS), 2011, pp. 2377-2380.